# An Overview of Integration of the Virtualization of Network Functions in the Context of Information Centric Networks

Antonio Cortés Castillo
Departamento de Informática
Universidad de Panamá
Panamá, República de Panamá
antonio.cortes@up.ac.pa

*Abstract*— Nowadays, the new network architectures with respect to the traditional network topologies must manage more data, which entails having increasingly robust and scalable network structures. As there is a process of growth, adaptability, and change in traditional data networks, faced with the management of large volumes of information, it is necessary to incorporate the virtualization of network functions in the context of information content networks, in such a way that there is a balance between the user and the provider at cost level and profit, the functions on the network. In turn, NFVs (Network Functions Virtualization) are considered network structures designed based on IT virtualization technologies, which allow virtualizing of the functions that can be found in the network nodes, which are connected through routing tables, which allows offering communication services for various types of customers. Therefore, information-centric networks (IC, unlike traditional data networks which proceed to exchange information between hosts using data packets and TCP/IP communication protocols, use the content of the data for this purpose where the data travels through the network is stored in a routing table located in the CR (Content Router) of the router temporality to be reused later, which allows for reducing operation costs and capital costs. The purpose of this work is to analyze how the virtualization of network functions is integrated into the field of information-centric networks. Also, the advantages and disadvantages of both architectures are considered and presented as a critical analysis when considering the current difficulties and future trends of both network topologies.

I. INTRODUCTION

Today, with the emergence of next-generation network architectures where IoT,5G,6G technologies are integrated, the massive data flow is constant and growing, causing changes at the hardware and software level in the network structure of the network in a dynamic and scalable way [1,2]. Therefore, the structures of traditional networks generate massive energy consumption, which implies high costs and large-scale investment in specialized labor that allows service providers to enter new network services to their customers [3]. At the same time, when these incompatibilities appear at the hardware level of the network architectures and the technological devices that interact in these traditional network structures, it makes network service providers go further than these traditional systems when using network structures virtualized standard to be deployed in the network. However, it must be considered that the use of network functions virtualization (NFV) allows the incorporation of new services for their customers in an agile and direct way, in which cloud service providers generate greater profits and scalability in the network architecture. Nevertheless, having a virtualization of the network functions, makes the physical devices decouple, allowing them to run logically and where the virtual machines continue to operate on standards servers. There are some network functions (NFs) related to firewalls, the DNS used to help virtualize network functions from a virtual perspective in the field of network functions virtualization, thus allowing the installation of virtual network functions anywhere and without investment in new equipment, which helps to optimize space and reduce power consumption of hardware level components [4]. However, the virtualization of network functions helps to reduce operating expenses and investment in new expenses, which allows the implementation of new services quickly and agile [4]. This means that NFVs maximize the use of virtualization techniques, which helps a scalable growth of the network architecture. Therefore, NFVs justify their use, since current networks that use the Internet are based on a TCP/IP communication model based on packet switching [5]. This type of point-to-point communication model, made up of full-duplex communication links between the network nodes, makes it unreliable and with a high redundancy rate in managing large volumes of information, underutilizing resources in a complex grid system. Given this situation, CSPs are testing new network technologies that allow the incorporation of new services, to existing ones, in ICN networks [6,7], which allows rapid recovery of the content that is distributed among multiple clients on the network securely and helping the network structure grow by achieving flexibility level at the network level. Therefore, the ICNs change the model in the way in which the nodes in a network communicate, since they go from packet switching to the management of content managed through the content in routers (CR), allowing cloud storage and virtualization of network functions to be successfully performed on the same fabric as content-centric networks. So, the new services that are going to be provided to customers are carried out in the ICNs, by means of the location of the service functions and the content management for the exchange of data [6,9].

The article is split into the following sections. In Section II., fundamental concepts of Information-Centric Networks versus Network Functions Virtualization are introduced. In section III., a critical review of the proposed architectures is performed. In

Section IV, the changes to take practice in the NFVs are discussed. Section V outlines the conclusions and future lines of research regarding ICN and NFV networks.

## II. Fundamental Concepts of Information-Centric Networks versus network Functions Virtualization

Customer demand for new services through state-of-the-art network architectures, for example, 4G, 5G, 6G, PON, are constantly in demand and growing, so today's Internet cannot satisfy all of them this demand, which demands new requirements at the level of network structure, both at the hardware and software level [10]. Some investigations, in the context of ICNs, analyze the positive contributions that NFVs provide in the field of ICNs, due to the limitations that the current Internet presents.

### A. Information-Centric Networks

Information-centric networks promise significant advances for what are next-generation networks in the realm of the Internet, allowing better support at the level of network, load balancing, and the management of big volumes of data [10,11]. In the context of ICNs, data storage in router containers (CR) is used, which allows a better distribution of data in the network, since this storage in routers is temporary using content addressing tables, thus allowing information to be shared among the clients that can be found in the various nodes [12]. It is also important to highlight the role of the content delivery network (CDN), which is made up of a conglomerate of distributed servers that store their content in router containers (CR) close to end users, allowing the fast transfer of files to be able to upload content on the Internet, by making use of HTML pages, JavaScript files, images, and videos, among others. Thus, various access points, which are identified as wireless routers, clients can access content stored on these routers, which are part of the ICN network infrastructure, which allows better performance and operation of the network, since the contents can be distributed among the users in the form of data packets, meeting the current and future needs of their clients, which helps to improve the current situation of the Internet [7,13]. The current network structure, the Internet, allows communication between nodes, using the full-duplex transmission medium, and in which, at the hardware level, computers play a key role, since through the domain name (DNS), these machines communicate with each other by taking and carrying out certain types of actions, for example, the allocation of IP addresses on the network. In the case of IPv6 addresses used at the device level, for example, mobile devices or whatever they may be, they must be able to change location arbitrarily, maintaining their respective existing connections. So, to attribute these capabilities to mobile devices, a mobile node is created to which a particular address is assigned in which it can be located, which allows the mobile node to be in a specific place and can use a specific address in that specific place but when the mobile node is out of the coverage range of the wireless network of the specific place, then and through the router, the message relay is carried out between the mobile node and the peripheral nodes with what it establishes communication. Meanwhile, the ICN network structures arise as a solution of how to get the data through the network and not how one machine connects to another (M2M). Some elements that make up information-centric networks are related to names, named data objects (NDOs), forwarding information, data routing, application programming, and temporary storage of content in routers. In the case of (NDOs), these must be unique, for each name object, since this avoids redundancy and duplication of unnecessary information, which helps safeguard the integrity and security of the data in space and time default [14]. At the same time, information-centric networks must manage the recovery of objects from the data that circulates on the network and guarantee information security by thus allowing efficient management of the distribution of user content on the Internet grid. Due to the above, the ICNs allow scalability and low-cost investments, unique identifiers for each object name, mobility among their users, and resilience in the network structure [15,35]. To guarantee security on the network, the ICNs use Transport Layer Security (TLS) [36], to guarantee the sending of information between their users at the client-server level, so establishing a level of trust between the client and the server is necessary to be able to deliver the information to their respective recipients in such a way that the information is not lost on its journey through the network. In this same direction, the ICNs filter unwanted information, for example, spam, thus avoiding the sending of unwanted information requests by customers, only responding to those requests that really need it [10]. This information filtering is performed at layer 3 and higher [16]. In this case, the ICNs use the contents temporarily stored in the routers, which allows generating more space to store other contents and thus attend to the requests of the users when they need it [17]. In the field of ICNs, a significant advance is the reduction in the delivery time of the contents to the users, which decreases the delay in the transmission of the data and increases the possibility of delivery of these contents to the mobile users [18,37]. Some investigations [19] propose that to overcome some of the weaknesses of the ICNs, Network Representation Learning (NRL) should be used. In Fig. 2., the fundamental elements of an information-centric network can be seen.

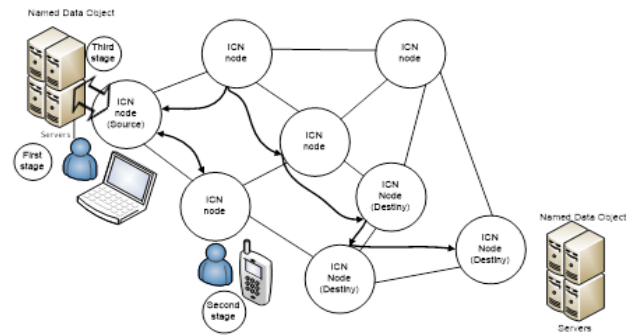

Fig. 2. Essential Elements of an Information-Centric Network

### B. Network Functions Virtualization

The virtualization of network functions allows for a more stable and optimized network structure at the level of the incorporation of resources and integration with heterogeneous and service network structures [20]. Thus, NFV integrate virtualization technologies and offer new alternatives to design, implement and manage network services [21]. Therefore, at the hardware level, NFVs divide the physical equipment from the functions to be subsequently executed, where reliability depends on functional and hardware capacity, respectively [22]. In turn, network functions virtualization transforms the fabric of the network, as virtualization enables the separation of software fabrics from hardware entities and the decoupling of location functions for faster replenishment of network services [21]. Thus, conglomerate network equipment, large scale servers, switches can be located for end user access, hosts used for data distribution and growing data centers.

Within the NFV context, one of the tasks to be carried out is segmentation at the software and hardware level since it allows a high degree of flexibility in the decentralization of network functions. Thus, the scalability processes that the network architecture allows to increase the flexibility at the level of the capacities and services in the network structure. Similarly, implementation and maintenance services can be carried out in the network topology more economically and dynamically [23]. In Fig.3., some differences in the approaches of traditional networks with respect to the paradigm of the virtualization of network functions can be observed.

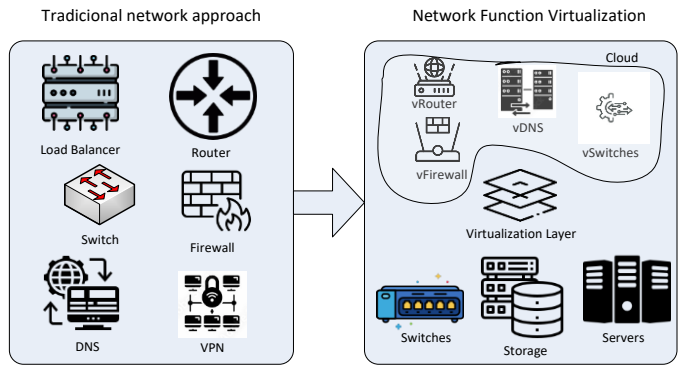

Fig. 3. Differences between the approaches of Traditional Networks and the virtualization of network functions

The compute instances that fulfill a specific function in the context of network functions virtualization (NFV) are:

- ***Virtual Machine Hypervisor:*** It is software used to establish virtualization from a predetermined communication layer level, which helps multiple operating systems to run one after the other while sharing the same physical computing resources. Using this software on a computer or physical servers allows the computer to divide its operating system and its applications from the hardware, thus providing savings in investment costs, agility and speed to the network, and a decrease in the downloading of applications.

- ***Containers:*** They are defined as executable software units in which the application code is a data package composed of libraries and dependencies that allows managing virtualization in a more agile and lightweight way, since it does not use a hypervisor, which allows it to faster access to resources and new applications in the cloud.

- ***Virtual Machine (MV):*** It is the virtual emulation or representation of a physical computer, in which the guests are represented through virtual machines and the hosts or nodes represent the physical machine. These virtual machines are carried out through virtualization, and each one has to its credit an operating system, and its respective computer applications.

In the case of virtual network functions (VNFs), run within virtual machines (VMs) or containers.

### III. CRITICAL REVIEW

#### A. Segmentation of the network to provide services to the clients in the ICNs in virtual environments

To provide better support at the level of performance and functionality to network structures segmentation of the network through virtualization techniques are used, since traditional networks that use IP addresses, between different communication nodes, are not sufficient to manage the load balance that it must tolerate the network at a certain time, in relation to the new emerging network architectures [12]. In this same direction, segmentation of the networks allows managing different types of speed at the network level to transmit data, high capacity in data transfers, wide coverage, and various types of security applied to the various applications that must be met for users to they can interact with you. The authors in [24] propose an ICN as a network structure that allows diversification at the level of the different data plans without having to make use of a physical network architecture. Being able to segment a physical network with various virtual environments, which are linked to various application services, by making use of the various functions offered by (NFV), allows to offer better performance and reliability of the network structure according to the services requested by the users [25]. Thus, a virtualization system (ICN) in the network architecture segmentation provides agile and reliable services, optimizing data flow in the network structure. Name-based networks, storage in routers, edge computing [26], security aspects [27,28], and mobility of users in ICNs are outlined in [24] and can be applied in the context of network architectures network. Similarly, being able to have a segmented network is achieved through the virtualization of the network structure, where users are allowed access to a variety of services, storage, technological resources, etc. In this same direction, the ICNs are emerging as a network structure that is made up of hardware and software components and that, by using network segmentation, makes it possible to offer a diversity of services in the context of virtual environments network architectures.

#### B. NFV network architectures

The network structures in the context of (NFV), having a series of functions that allow network virtualization, host a series of services using the features at the hardware level of the network where clients can access a series higher quality service through cloud service providers (CSPs) [29]. Therefore, in the context of NFV, most of the functions that make up this type of network architecture at the virtual level are executed in software mode in the network nodes located in virtual machines [30]. In the same way, the virtualization of the functions allows the execution of various data plans, running and accessing various applications, and access to the functions in the control plane layer at the level of fixed and mobile data network telephony. At the same time, there are 4 types of network architectures that constitute the virtualization of network functions and that we detail below.

- Centralized Network Structures

In the field of NFV, this type of network structure is the most used due to its simplicity and generalities, used by cloud service providers (CSPs) for the construction of virtualization of network functions. In the case of VNFs, they are installed in data centers (DCs) or in the network infrastructures of cloud service providers. Due to the above, VNFs are built inside data centers (DCs). The above allows access to virtualized network functions through an Ethernet connection using DNS located on centralized servers used in the construction of the VNF ecosystems. In Fig. 4., the structure of the centralized network in the field of NFVs is observed.

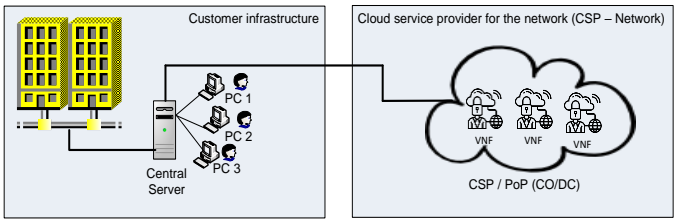

Fig. 4. Network architecture centralizes in NFV

- Decentralized Network Architectures

In the context of decentralized NFV network structures, the virtualized network functions are localizable in the customer's infrastructure. Thus, there are no VNFs that can be reached in the data centers (DCs). In Fig.5., the decentralized network architecture of a NFV can be seen.

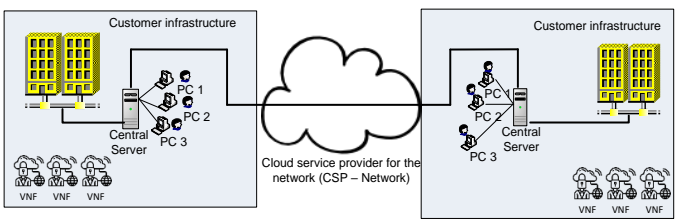

Fig.5. Decentralized network architecture in the NFV

- Distributed Network Architectures

The distributed network structure in the NFV context is a combination of the centralized and decentralized network structures. Therefore, virtualized network functions are in the data centers (DCs) and the customer's infrastructure. In turn, the virtualized network functions are randomly grouped, configured, and bundled together in a distributed network infrastructure. In the distributed network structure, the growth, integrity, and viability in the construction of the VNFs are optimized since it allows to improve the load balance and the data flow in the network. In Fig. 6., the structure of the distributed network in an NFV is observed.

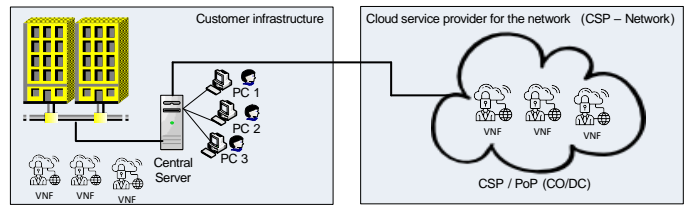

Fig. 6. Distributed network architecture in the NFV

- Edge network architecture for Dynamic Virtual Network (DVNe)

In edge network architectures for dynamic virtual networks, the virtualization of services offered to users is performed at the edge of the network in the virtual equipment that is in the customer infrastructure (vCPE). Thus, cloud computing is integrated into the NFV network structure. In addition, the virtual network functions, apart from being distributed in the cloud service provider´s data center and customer infrastructure, are also expanded at the cloud edge. In Fig. 7., the edge network architecture for Dynamic Virtual Networks (DVNe) can be seen.

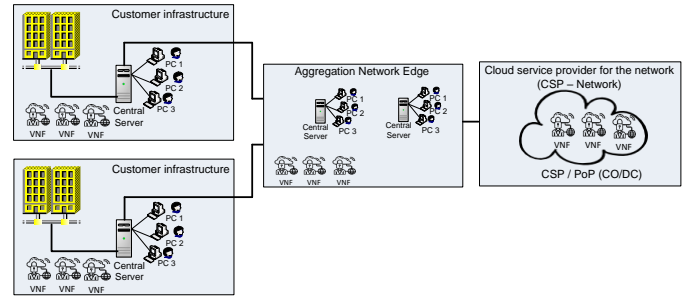

Fig.7. Edge network architecture for dynamic virtual networks (DVNe) in NFV

## C. Integrating NFVs into ICNs

If we take into consideration that at the level of the NFVs in the virtualization layer these are made up of routers, firewalls, switches and virtual DNS, the link with the ICNs would be carried out through the contents of the routers using the Named Data Object (NDO) that would be the equivalent to the virtual functions of the NFVs. In Fig. 8., we can see this integration of the NFVs in the ICNs.

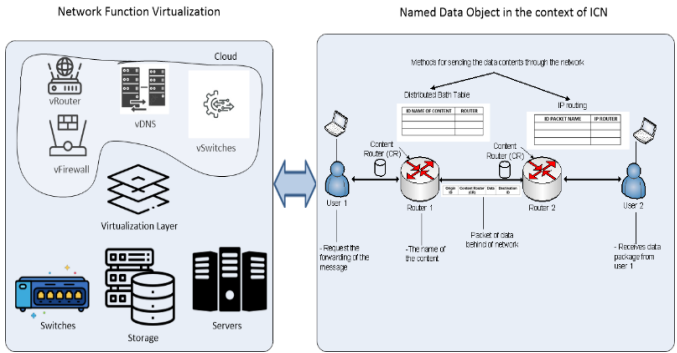

Fig.8. Integrating NFVs into ICNs

Therefore, each one of the dynamic tables assigned to each one of the routers, which through the (NDO) distribute the information in the network through IP addresses, the collaborative process of integration between both network technologies, virtualization of network functions and information content networks would be carried out in an optimal and cost-effective manner, both for users and providers.

## IV. CHANGE TO TAKE PRACTICE IN NFV

In the field of visualizing network functions, the benefits that these network architectures offer their clients in the context of TIC is still expanding, which means that the services that are going to be offered from a virtual perspective using a variety of functions through the network, is still in its early stages. In these circumstances, the virtualization of network functions must still mature at the technological level, since there are still aspects that must be tested and investigated, especially from a practical perspective in which solutions that can be applied in this context emerge. In the same way, the virtualization of network functions has been a great challenge for current management systems, since if it is not managed correctly, this can generate a malfunction of the network when providing new services on the network to the respective users and it may be that some functions that previously worked on the servers now do not work properly [23].

In this same direction, an optimal level of service per user must be guaranteed and that the functions that are going to be required can be used without any inconvenience, since if a problem arises it must be resolved as soon as possible, in such a way that the offered service continues to be provided satisfactorily.

However, NFVs must face new challenges related to commercial implementation in the Telecommunications market, because there are other products that compete with them. According to a study by The IHS Market 2017 Carrier NFV Strategies [31], the fact that this type of technology is not yet fully mature or lacks implementation of new developments. It could also be because there is an NFV integration issue with current network topologies. In the same way, the lack of knowledge and the use of standards with their respective regulations that support the use of NFV technologies are challenges that must be faced. Despite the inconveniences that may arise, it is possible to use and implement best practices, such as learning and experimenting through training that helps to understand the processes and technologies of NFV. In these training processes, use can be made of open software tools aimed at understanding NFV, since they are open software tools, not blocked, and to which users can access.

At the same time, NFVs are more oriented towards remote programming using network resources at the function level, which can activate some type of malicious software that causes some type of damage in the NFV environment. The foregoing indicates that security in the field of VNFs is not fully developed [32], turning out to be a challenge, which can be eliminated and removed from that malicious intruder that can be found in the network system. Other challenges to overcome are related to improving NFV performance, since high-performance processing must be achieved in data packets that are sent over the network in a virtualized environment. Similarly, this high-performance processing optimization involves the use of software and networks in a virtualized network structure [33], generating an excess load balance of data in the network. Therefore, there must be a balance in the network system, as well as the use of network functions in the network.

In this same direction, other challenges to overcome is the measurement of network performance, between nodes that make up the network structure using shared processors and TCP/UDP communication protocols, the latter turning out to be very unstable because they are protocols that use IP addresses and in the case of UDP, it is a connectionless protocol, in which the data packets travel in any direction in the network, which causes poor performance in the network structure in it network virtualization framework. One of the ways to improve performance is the use of connection-oriented protocols and the use of containers on routers that are used for data transmission. On the other hand, it is necessary to consider aspects related to the security of the network, since it must be well protected against unwanted intruders, especially in those cases in which a new NFV is being implemented. In the latter case, data drift in a virtual machine system (VMS) located in a data center must be considered in the context of VNFs, where manual server-level management of virtual machine systems it becomes impossible to do. Some providers help solve this type of situation by using a Firewall that provides security to the entire network, made up of virtualized equipment and network functions, such that throughput on the network is in equilibrium.

## V. CONCLUSIONS AND FUTURE WORK

The NFV network structures allow you to provide benefits to your customers and cloud service providers by allowing them to grow in modern network infrastructure. Despite the challenges faced by the virtualization of network functions, these benefits make it possible to incorporate through innovation those aspects that are necessary to mature, minimizing latent threats in the context of security in the context of networks. These challenges to be solved, using technical aspects provided by the telecommunications industry, by combining experience and technological resources, are complemented by collaborative processes, where valuable information is shared in the NFV field, in terms of standards and regulations in this type of network architecture. The achievement of agreements in terms of standards makes it possible to address new technical challenges that could be faced in the medium term and provide solutions that help to launch new virtualized services for mass consumption for users. The above requires significant transformations and great efforts by cloud service providers, which would help the transformation of computing at the edge. Therefore, through automation, organization, and virtualization, it would be possible to obtain improvements in the operation of the NFV, which would transform the current network infrastructures in terms of implementation, operation, and management of large volumes of data.

At the same time, to seamlessly migrate services from one network infrastructure to another, there must be planning and management of various providers.

Likewise, NFV has gradually become a network technology that can have a great innovative impact on traditional and next-generation network structures. Thus, because it is a growing technology, it is still pending to explore all its potential, and challenges or difficulties that it may present to obtain a variety of benefits that this network technology offers.

As future work, it would be necessary to investigate the integration of NFV with other next-generation network technologies as stated in [34], where Fat Tree or BCube network architectures are mentioned in the context of data centers.